\documentclass[traditabstract]{aa}
\usepackage{txfonts}
\usepackage{graphicx}
\usepackage{natbib}
\bibpunct{(}{)}{;}{a}{}{,}

\begin{document}

\title{Period decrease in three SuperWASP eclipsing binary candidates\\ near the short-period limit}
\author{M.~E.~Lohr\inst{\ref{inst1}}\and
  A.~J.~Norton\inst{\ref{inst1}}\and U.~C.~Kolb\inst{\ref{inst1}}\and D.~R.~Anderson\inst{\ref{inst2}}\and F.~Faedi\inst{\ref{inst3}}\and R.~G.~West\inst{\ref{inst4}}}
\institute{Department of Physical Sciences, The Open University,
  Walton Hall, Milton Keynes MK7\,6AA, UK\\ \email{Marcus.Lohr@open.ac.uk}\label{inst1}\and Astrophysics Group, Keele University, Staffordshire ST5\,5BG, UK\label{inst2}\and Astrophysics Research Centre, Main Physics Building, School of Mathematics \& Physics, Queen's University, University Road, Belfast BT7\,1NN, UK\label{inst3}\and Department of Physics and Astronomy, University of Leicester, Leicester LE1\,7RH, UK\label{inst4}}
\date{Received / Accepted}

\abstract {SuperWASP light curves for 53 W\,UMa-type eclipsing binary
(EB) candidates, identified in previous work as being close to the
contact binary short-period limit, were studied for evidence of period
change.  The orbital periods of most of the stars were confirmed, and
period decrease, significant at more than 5~$\sigma$, was observed in
three objects: \object{1SWASP\,J174310.98+432709.6}
($-0.055\pm0.003$~s~yr\textsuperscript{-1}),
\object{1SWASP\,J133105.91+121538.0}
($-0.075\pm0.013$~s~yr\textsuperscript{-1}) and
\object{1SWASP\,J234401.81$-$212229.1}
($-0.313\pm0.019$~s~yr\textsuperscript{-1}).  The magnitudes of the
observed period changes cannot be explained by magnetic braking or
gravitational radiation effects, and are most likely primarily due to
unstable mass transfer from primary to secondary components, possibly
accompanied by unstable mass and angular momentum loss from the
systems.  If these period decreases persist, the systems could merge
on a relatively short timescale.}

\keywords{stars: variables: general - binaries: close - binaries: eclipsing - stars: individual: V1067\,Her - stars: individual: \mbox{GSC\,2314$-$0530}}
\titlerunning{Period decrease in three SuperWASP eclipsing binary candidates}
\authorrunning{M.~E.~Lohr et al.}

\maketitle
\section{Introduction}

The SuperWASP project has been surveying bright ($V\sim$8--15
mag) stars across almost the whole sky since 2004, looking for
photometric variations indicative of exoplanetary transits
\citep{pollacco}.  However, in addition to 67 exo\-planets announced
by early February 2012\footnote{http://exoplanet.eu/catalog.php}, it
has discovered several tens of thousands of new variable stars, the
majority of which appear to be eclipsing binaries (EB) (Payne in prep.).
The cameras' observational cadence permits detection of periods from
several months down to about 20 minutes (e.g. pulsations in Am stars
measured by \citealt{smalley}).  Hence the SuperWASP archive
represents a useful source of evidence for short-period EBs.

W\,UMa-type variables, which usually represent contact binaries,
exhibit a fairly sharp lower limit to their observed orbital periods,
of around 0.22~d \citep{ruc07}.  Suggested explanations for this
cut-off have involved the stars reaching full convective limit
\citep{ruc92} or an effective low mass limit, given the finite age of
the Universe, on the assumption that angular momentum loss rates are
related to stellar masses \citep{stepien}.  However, the convective
limit would not provide a full explanation \citep{ruc92}, and
Stepien's low mass limit does not appear to hold true for some
recently-discovered systems \citep{jiang}
e.g. \object{\mbox{GSC\,2314$-$0530}}.  Hence \citeauthor{jiang} have
suggested that dynamic mass transfer instability develops in detached
binary systems possessing both low initial primary mass and a low mass
ratio, once the primary fills its Roche lobe, leading to a rapid
merger of the binary.  Stable mass transfer would however lead to
formation of a contact binary at the point of Roche lobe overflow,
explaining the short-period limit for \emph{contact} binaries with
appropriate primary mass and mass ratio.

\citet{norton} presented evidence for 53 candidate W\,UMa-type EBs
observed using SuperWASP, with periods close to the short-period
limit, of which 48 were new discoveries.  Some of these, when folded
at their determined period, failed to give well-defined light curves,
which could be the result of period variability.  Therefore our aim
here was to determine whether any of the 53 candidate EBs showed
evidence of period changes during their observation by SuperWASP.
Period decreases might indicate a detached or semi-detached binary
undergoing unstable mass transfer likely to end in merger, or even a
contact binary approaching merger (as observed by \citealt{tylenda}).

\section{Method}

The best orbital period values for the 53 stars were checked using a
three-step custom-written IDL procedure.  First, an approximation to
the half-period was found by fitting a sinusoidal function, using the
Levenberg-Marquardt algorithm \citep{lev,marq}, to light curve data
from three optimal nights of observation.  The inverse-variance
weighted average half-period, and its uncertainties, were then used to
constrain the search range for a Lomb-Scargle periodogram
\citep{lomb,scargle,hornebal} to determine a more precise estimate for
the half-period, which was doubled to give an estimate for the orbital
period.  Finally, the light curve data set was folded on trial periods
within a narrow range, and the dispersion of flux values within phase
bins was minimized to find the optimum period to the nearest 0.01~s.

A reference light curve minimum was then automatically selected as
near to the middle of the data set as possible (since the period here
would be expected to approximate the `average' period found for the
whole data set in cases where the period was changing linearly).  The
time of zero-point minimum $t_{\mathrm{0}}$ was more precisely found
by fitting a sinusoidal function to nearby points; the optimum period
length $P_{\mathrm{calc}}$ was then used to calculate expected times
of minimum light ($C = t_{\mathrm{0}}+P_{\mathrm{calc}}E$, where
$E$ is the period number, or epoch, relative to $t_{\mathrm{0}}$)
within the full time-range observed.  Actual times of minimum $O$ near
these values were determined using quadratic fitting (which proved
more robust than Gaussian or sinusoidal fits), with poor fits and
inadequately-sampled minima being automatically rejected.

In this way a series of observed minus calculated (O$-$C) values were
obtained, and plotted against epoch.  A second series was found for
the other set of light curve minima (since EBs exhibit a primary and
secondary eclipse during each orbital period).  Such O$-$C diagrams
will be expected to follow a quadratic curve if the period is changing
linearly, allowing the determination of the rate of change of period
with respect to epoch.  Since $O = t_{\mathrm{0}}+P(E)E$ when
$P = P(E)$, we have:
\begin{equation}
O-C = P(E)E-P_{\mathrm{calc}}E = aE^2+bE+c,
\end{equation}
so, differentiating with respect to $E$:
\begin{equation}
\frac{\mathrm{d}P}{\mathrm{d}E}E+(P(E)-P_{\mathrm{calc}})=2aE+b,
\end{equation}
giving $\mathrm{d}P/\mathrm{d}E=2a$.  Therefore an attempt was
made to fit a quadratic function to each O$-$C series, from which a
value for the rate of period change (per period) could be determined.
The inverse-variance weighted average period change (across both O$-$C
series) was thus found for each star, and could be converted into an
approximate rate of change $\mathrm{d}P/\mathrm{d}t$ in
s~yr\textsuperscript{-1}.

The validity of the method and code described above was tested using
several synthetic data sets.  Even with noisy data, it proved able to
recover input periods correct to 6 to 7 significant figures (s.f.),
and input linear period changes correct to
0.1~s~yr\textsuperscript{-1}, significant at over 7~$\sigma$.  The
latter error is largely due to the approximation of a continuous
function of time by a discontinuous function of epoch in the customary
differential equation as pointed out by \citet{kopkurth}.

\section{Results}

\begin{figure}
\resizebox{\hsize}{!}{\includegraphics{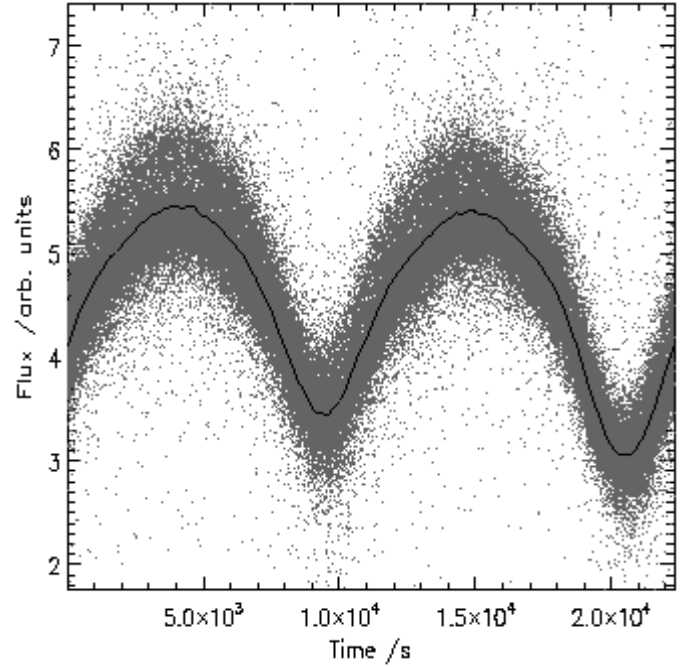}}
\caption{Light curve for object J1743, folded at period of 22300.517~s, with mean values for binned data overplotted.}
\label{star17lc}      
\end{figure}

\begin{figure}
\resizebox{\hsize}{!}{\includegraphics{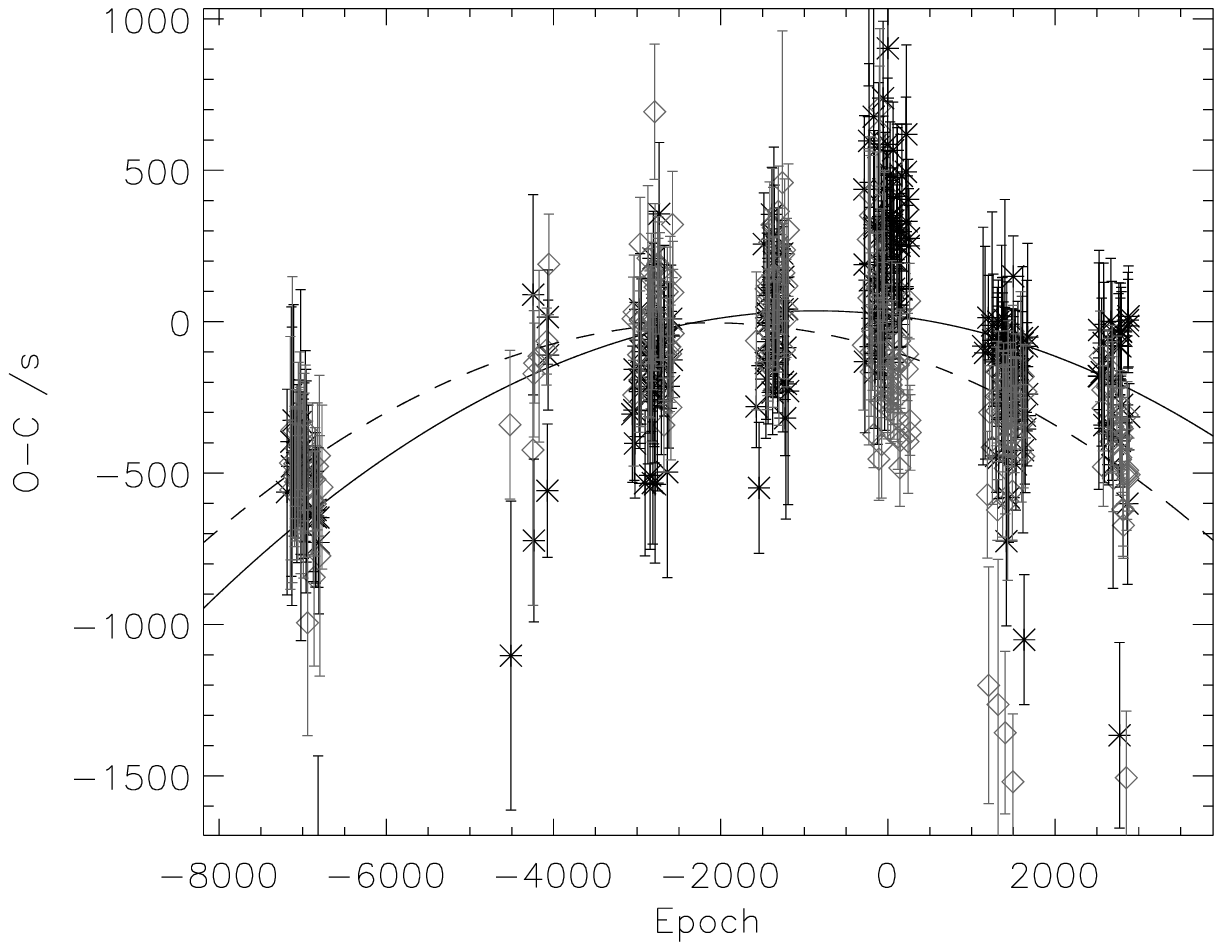}}
\caption{O$-$C diagram for object J1743, with best fit quadratic curves overplotted.  Black stars and solid line indicate primary eclipse series; grey open diamonds and dashed line indicate secondary eclipse series.  Period decrease significant at 21~$\sigma$ is indicated.}
\label{star17oc}      
\end{figure}

\begin{figure}
\resizebox{\hsize}{!}{\includegraphics{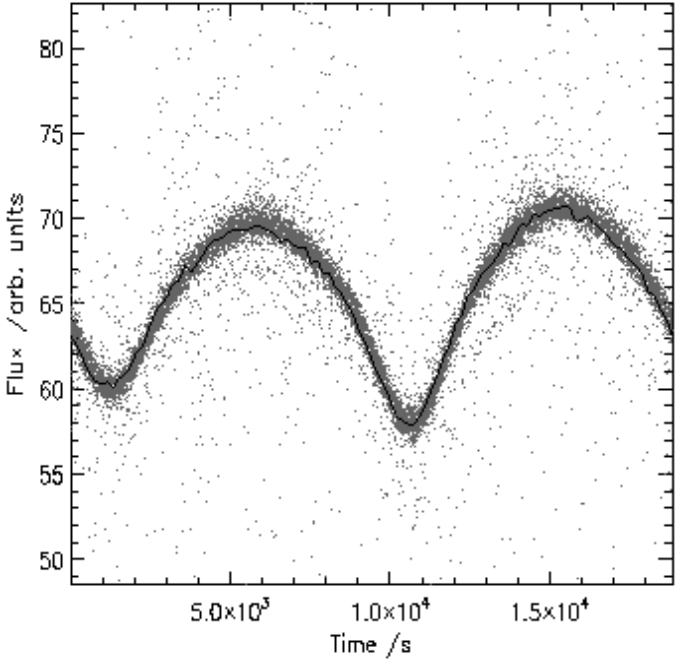}}
\caption{Light curve for object J1331, folded at period of 18836.380~s, with mean values for binned data overplotted.}
\label{star41lc}      
\end{figure}

\begin{figure}
\resizebox{\hsize}{!}{\includegraphics{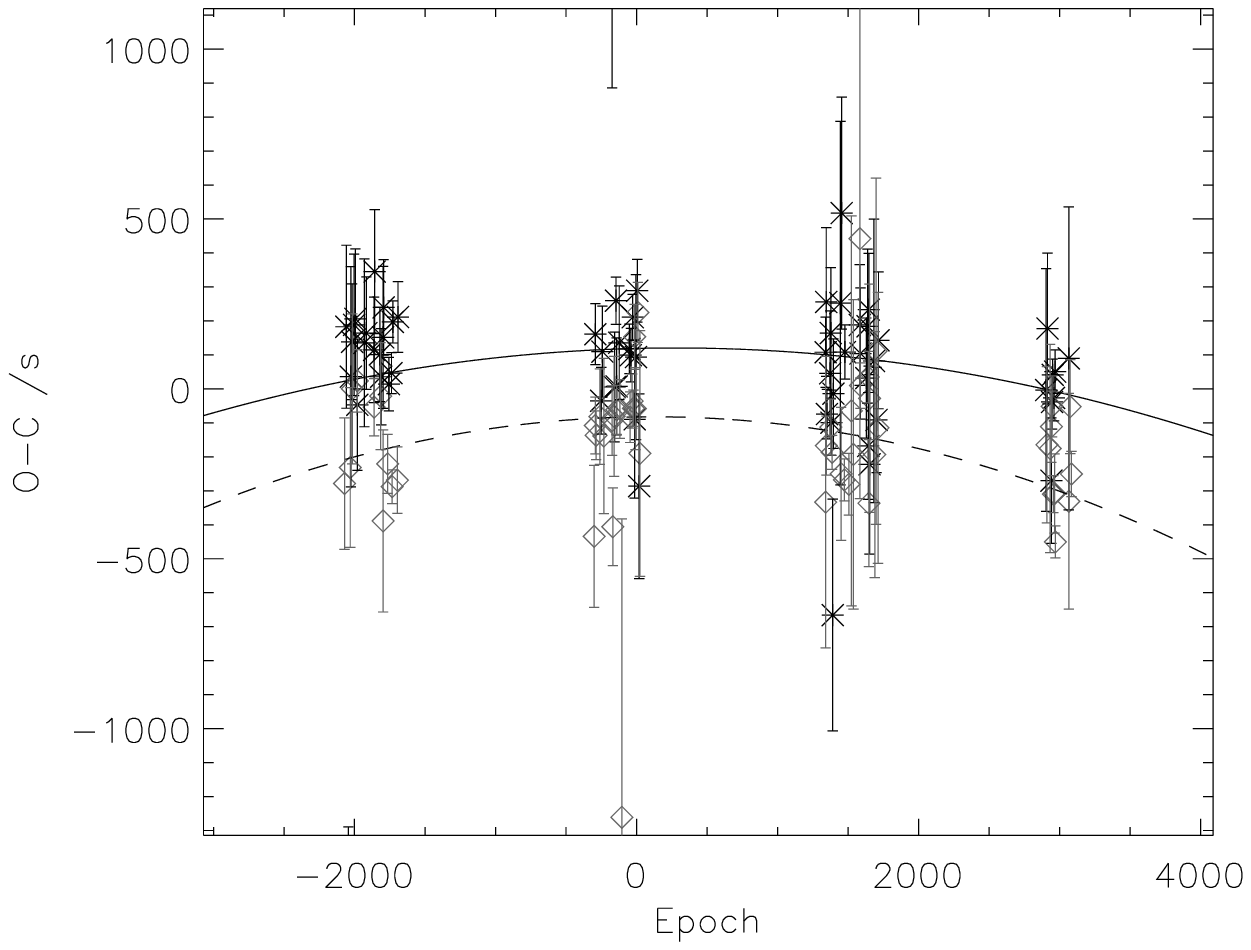}}
\caption{O$-$C diagram for object J1331, with best fit quadratic curves overplotted.  Black stars and solid line indicate primary eclipse series; grey open diamonds and dashed line indicate secondary eclipse series.  Period decrease significant at 5~$\sigma$ is indicated.}
\label{star41oc}      
\end{figure}

\begin{figure}
\resizebox{\hsize}{!}{\includegraphics{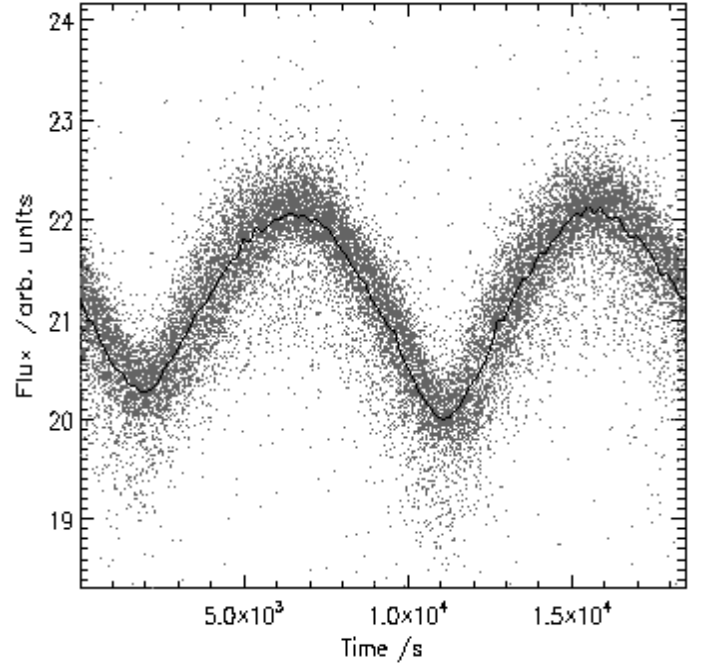}}
\caption{Light curve for object J2344, folded at period of 18461.591~s, with mean values for binned data overplotted.}
\label{star44lc}      
\end{figure}

\begin{figure}
\resizebox{\hsize}{!}{\includegraphics{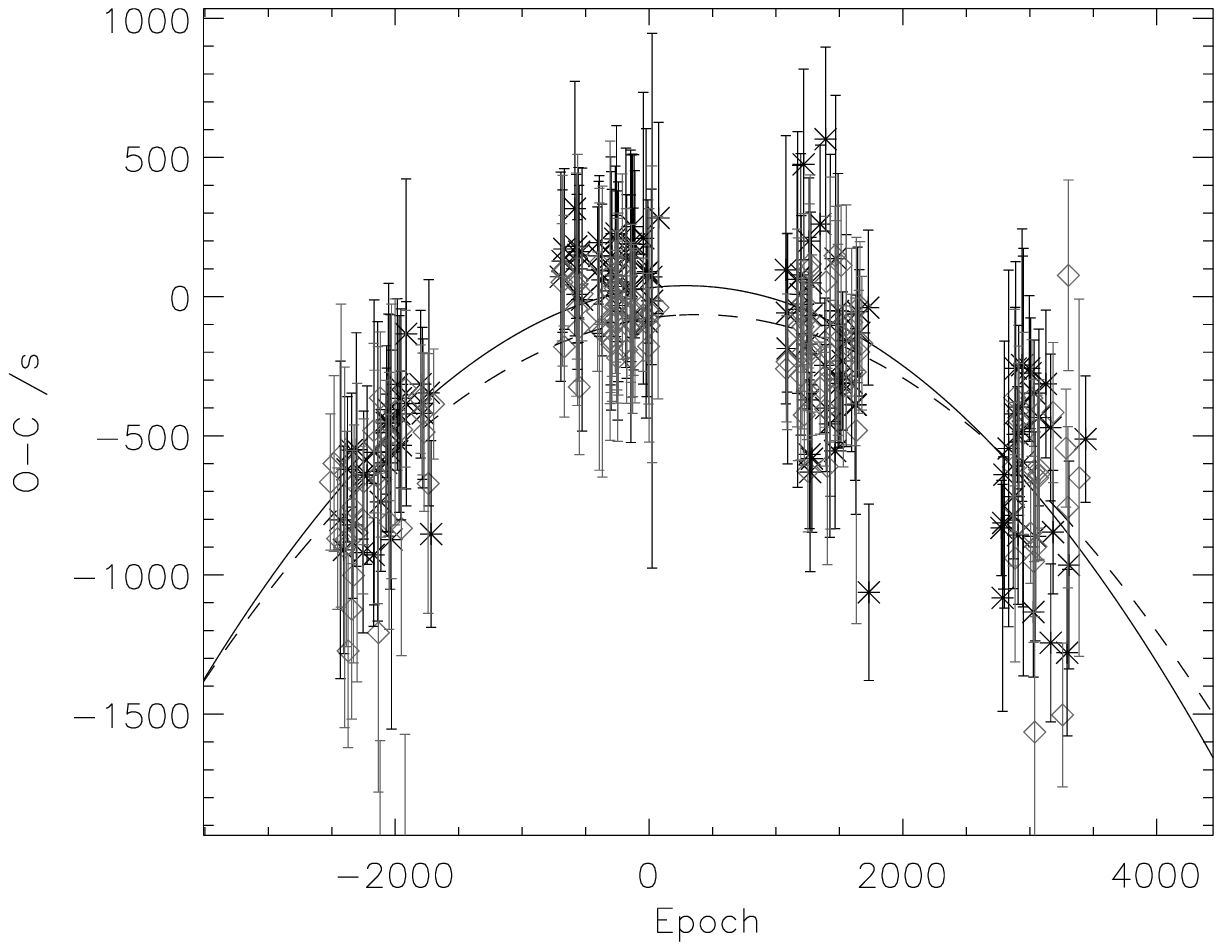}}
\caption{O$-$C diagram for object J2344, with best fit quadratic curves overplotted.  Black stars and solid line indicate primary eclipse series; grey open diamonds and dashed line indicate secondary eclipse series.  Period decrease significant at 16~$\sigma$ is indicated.}
\label{star44oc}      
\end{figure}

\begin{figure}
\resizebox{\hsize}{!}{\includegraphics{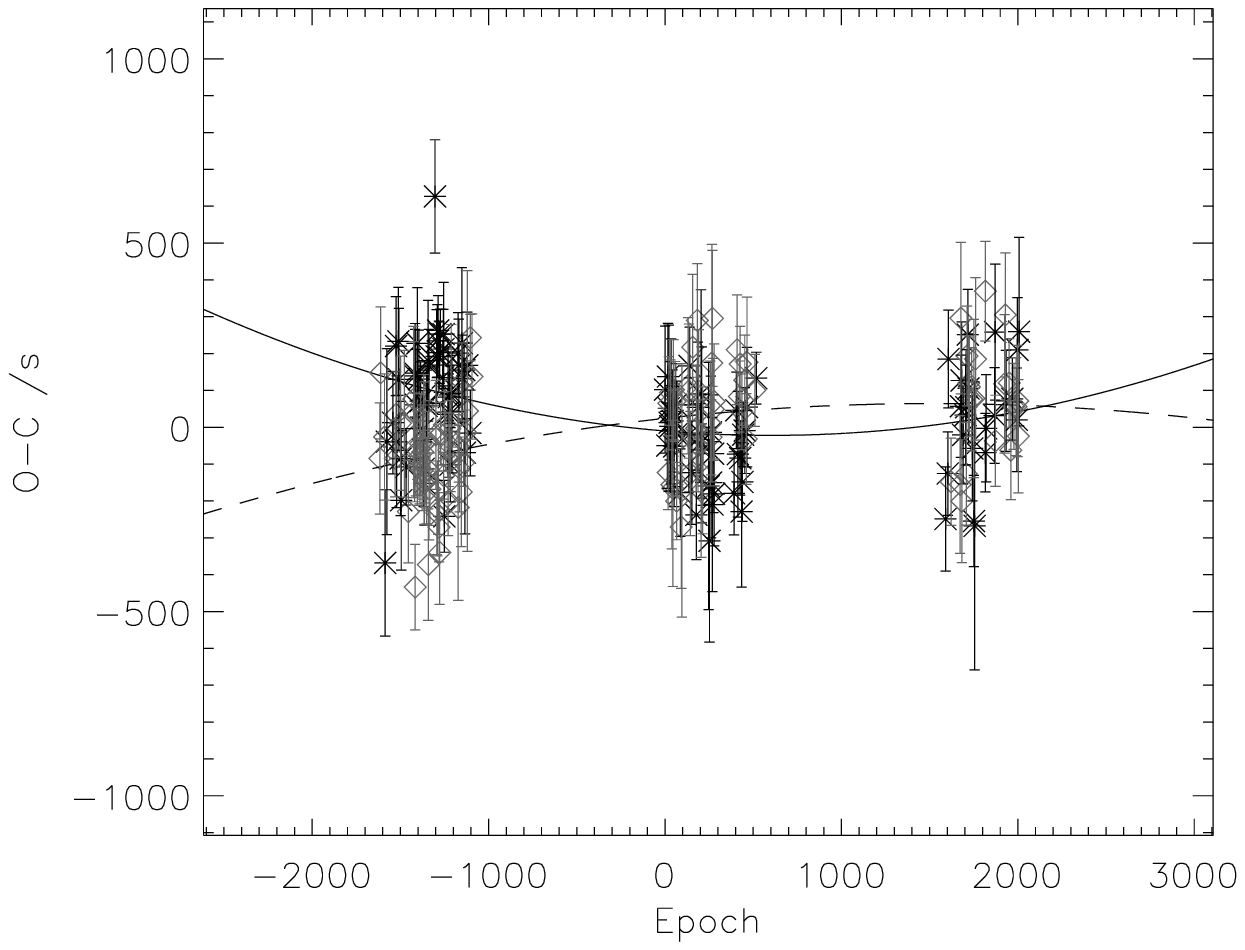}}
\caption{O$-$C diagram for object J0509, with best fit quadratic curves overplotted.  Black stars and solid line indicate primary eclipse series; grey open diamonds and dashed line indicate secondary eclipse series.  No significant period change is indicated.}
\label{star13oc}      
\end{figure}

The optimum orbital periods found for the 53 stars are given in
Table~\ref{53data}.  In the majority of cases, identical periods (to 4
or 5~s.f.) to those found in \citet{norton} were obtained.  Object
J0554's period was given incorrectly in table 1 of
\citeauthor{norton}, but correctly in the appendix.  For nine objects,
a period approximately $1/8$ longer was obtained by this method;
manual checks indicated that the longer periods found here were more
plausible (e.g. the folded light curve was more clearly defined; the
primary and secondary minima exhibited more difference in depths).  In
the earlier paper, only candidate periods below 20000~s
($\sim$~0.2315~d) had been investigated, allowing these longer periods
to be missed.  Therefore the 53 stars probably do not all possess
periods below 20000~s, but since all are below about 22600~s
($\sim$~0.2616~d), they still constitute a useful sample of notably
short-period EB candidates.

\begin{table*}
\caption{Period results for 53 SuperWASP eclipsing binaries.}
\label{53data}
\centering
\begin{tabular}{l l l l l l l l l}
\hline\hline
SuperWASP ID & Original & Short & $P$ & $\delta P$ & $\dot{P}$ & $\delta \dot{P}$ & $\frac {\dot{P}} {P}$ & Significance \\
(Jhhmmss.ss$\pm$ddmmss.s) & Number\tablefootmark{a} & Name & (s) & (s) & (s~yr\textsuperscript{-1}) & (s~yr\textsuperscript{-1}) & (yr\textsuperscript{-1}) & ($\sigma$) \\
\hline
J000437.82+033301.2 & 3 & J0004 & 22593.528\tablefootmark{b} & 0.009 & +0.095 & 0.087 & $+4.2\times10^{-6}$ & 1 \\
J003033.05+574347.6 & 27 & J0030 & 19580.280 & 0.004 &  &  &  &  \\
J004050.63+071613.9 & 15 & J0040 & 19809.204 & 0.004 &  &  &  &  \\
J022050.85+332047.6\tablefootmark{c} & 53 & J0220 & 16643.642 & 0.009 &  &  &  &  \\
J022727.03+115641.7 & 47 & J0227 & 18226.361 & 0.009 &  &  &  &  \\
J030749.87$-$365201.7 & 26 & J0307 & 19584.395 & 0.009 &  &  &  &  \\
J034439.97+030425.5 & 11 & J0344 & 19861.395 & 0.009 &  &  &  &  \\
J040615.79$-$425002.3 & 34 & J0406 & 19209.980 & 0.009 & $-0.029$ & 0.026 & $-1.5\times10^{-6}$ & 1 \\
J041120.40$-$230232.3 & 43 & J0411 & 18690.406 & 0.009 &  &  &  &  \\
J041655.13$-$492709.8 & 5 & J0416 & 19959.767 & 0.009 &  &  &  &  \\
J044132.96+440613.7 & 19 & J0441 & 19712.787 & 0.006 &  &  &  &  \\
J050904.45$-$074144.4 & 13 & J0509 & 19835.305 & 0.009 &  &  &  &  \\
J051501.18$-$021948.7 & 6 & J0515 & 19950.121 & 0.009 &  &  &  &  \\
J052036.84+030402.1 & 1 & J0520 & 19993.301 & 0.009 & $-0.037$ & 0.025 & $-1.8\times10^{-6}$ & 1 \\
J055418.43+442549.8 & 39 & J0554 & 18878.004 & 0.003 &  &  &  &  \\
J064501.21+342154.9 & 37 & J0645 & 21480.415\tablefootmark{b} & 0.009 &  &  &  &  \\
J074658.62+224448.5 & 38 & J0746 & 19081.403 & 0.005 &  &  &  &  \\
J080150.03+471433.8 & 42 & J0801 & 18793.167 & 0.007 &  &  &  &  \\
J092328.76+435044.8 & 50 & J0923 & 20292.982\tablefootmark{b} & 0.004 & +1.90 & 0.95 & $+9.37\times10^{-5}$ & 2 \\
J092756.25$-$391119.2 & 30 & J0927 & 19469.696 & 0.009 & $-0.10$ & 0.05 & $-5.1\times10^{-6}$ & 1 \\
J093012.84+533859.6 & 22 & J0930 & 19674.550 & 0.001 & (+1.40) & (0.13) & ($+7.11\times10^{-5}$) & (10)\tablefootmark{d} \\
J111931.48$-$395048.2\tablefootmark{e} & 14 & J1119 & 19827.651 & 0.007 & +0.033 & 0.014 & $+1.7\times10^{-6}$ & 2 \\
J114929.22$-$423049.0 & 23 & J1149 & 19639.554 & 0.009 &  &  &  &  \\
J115557.80+072010.8 & 25 & J1155 & 19614.222 & 0.009 & &  &  &  \\
J115605.88$-$091300.5 & 48 & J1156 & 18222.626 & 0.005 &  &  &  &  \\
J120110.98$-$220210.8 & 24 & J1201 & 19627.832 & 0.009 &  $-0.21$ & 0.09 & $-1.1\times10^{-5}$ & 2 \\
J121906.35$-$240056.9 & 29 & J1219 & 19558.186 & 0.009 &  &  &  &  \\
J130920.49$-$340919.9 & 33 & J1309 & 19253.528 & 0.009 &  &  &  &  \\
J133105.91+121538.0 & 41 & J1331 & 18836.380 & 0.004 & $-0.075$ & 0.013 & $-4.0\times10^{-6}$ & 5 \\
J142312.63$-$222425.1 & 51 & J1423 & 18112.792 & 0.009 &  &  &  &  \\
J150822.80$-$054236.9 & 10 & J1508 & 22469.214\tablefootmark{b} & 0.007 & $-0.026$ & 0.016 & $-1.2\times10^{-6}$ & 1 \\
J151652.90+004835.8 & 49 & J1516 & 18207.370 & 0.009 &  &  &  &  \\
J155822.10$-$025604.8 & 9 & J1558 & 22470.705\tablefootmark{b} & 0.009 & $-0.19$ & 0.13 & $-8.3\times10^{-6}$ & 1 \\
J160156.04+202821.6 & 28 & J1601 & 19572.134 & 0.009 & +0.12 & 0.05 & $+6.0\times10^{-6}$ & 2 \\
J161334.28$-$284706.7 & 12 & J1613 & 19852.817 & 0.004 & $-0.063$ & 0.055 & $-3.2\times10^{-6}$ & 1 \\
J170240.07+151123.5\tablefootmark{f} & 4 & J1702 & 22589.758\tablefootmark{b} & 0.004 & &  &  &  \\
J173003.21+344509.4 & 32 & J1730 & 19328.922 & 0.009 & &  &  &  \\
J173828.46+111150.2 & 35 & J1738 & 21546.731\tablefootmark{b} & 0.009 & \tablefootmark{g} &  & & \\
J174310.98+432709.6\tablefootmark{h} & 17 & J1743 & 22300.517\tablefootmark{b} & 0.009 & $-0.0546$ & 0.0025 & $-2.45\times10^{-6}$ & 21 \\
J180947.64+490255.0\tablefootmark{i} & 20 & J1809 & 19688.490 & 0.009 & +0.028 & 0.010 & $+1.4\times10^{-6}$ & 2 \\
J183738.17+402427.2 & 36 & J1837 & 19121.227 & 0.009 &  &  &  &  \\
J195900.31$-$252723.1 & 46 & J1959 & 20575.395\tablefootmark{b} & 0.009 &  &  &  &  \\
J210318.76+021002.2 & 16 & J2103 & 19750.199 & 0.009 & &  &  &  \\
J210423.76+073140.4 & 52 & J2104 & 18065.091 & 0.009 &  &  &  &  \\
J212454.61+203030.8 & 21 & J2124 & 19684.790 & 0.009 & $-0.41$ & 0.29 & $-2.1\times10^{-5}$ & 1 \\
J212808.86+151622.0 & 31 & J2128 & 19426.310 & 0.009 &  &  &  &  \\
J214510.25$-$494401.1 & 18 & J2145 & 19712.951 & 0.009 &  &  &  &  \\
J220734.47+265528.6 & 2 & J2207 & 19978.751 & 0.009 &  &  &  &  \\
J221058.82+251123.4 & 45 & J2210 & 18402.957 & 0.009 &  &  &  &  \\
J224747.20$-$351849.3 & 40 & J2247 & 18853.775 & 0.009 & +0.77 & 0.64 & $+4.1\times10^{-5}$ & 1 \\
J232607.07$-$294130.7 & 8 & J2326 & 19882.132 & 0.004 &  &  &  &  \\
J234401.81$-$212229.1 & 44 & J2344 & 18461.591 & 0.007 & $-0.313$ & 0.019 & $-1.70\times10^{-5}$ & 16 \\
J235333.60+455245.8 & 7 & J2353 & 19936.123 & 0.009 &  &  &  &  \\
\hline
\end{tabular}
\tablefoot{
Period changes included only where significance is at least 1$\sigma$.
\tablefoottext{a}{In \citet{norton}.}
\tablefoottext{b}{Period found here is longer than in \citet{norton}.}
\tablefoottext{c}{GSC\,2314$-$0530.}
\tablefoottext{d}{Data contaminated by nearby star's light curve.}
\tablefoottext{e}{ASAS\,J111932$-$3950.8.}
\tablefoottext{f}{ROTSEI\,J170240.11+151122.7.}
\tablefoottext{g}{Insufficient data to calculate both O$-$C series.}
\tablefoottext{h}{V1067\,Her.}
\tablefoottext{i}{V1104\,Her.}
}
\end{table*}

Table~\ref{53data} also gives the period change results.  Object J1738
only had enough minima data for a single O$-$C series, so was
excluded.  Object J0930 appeared to show significant period increase,
but its folded light curve and O$-$C diagram suggested an anomaly in
the data, and a check of the sky within 1$\arcmin$ of the object
revealed a nearby bright star which appears to have contaminated the
light curve; this object was also therefore excluded at this stage.
Of the remaining 51 objects, 17 showed period change inconsistent with
zero, at 1~$\sigma$ confidence, with 6 cases of apparent period
increase and 11 of decrease.  However, given the number of objects
studied, we might expect approximately this number of
marginally-significant results \emph{purely by chance}, and so we do not
claim that all these 17 results indicate genuine period changes
(although this is a possibility).

More notable is that three objects show a period change significant at
more than 3~$\sigma$, where we would expect no significant results by
chance alone in a set of 51 objects.  These three stars underwent
significant decreases in period during the time they were observed by
SuperWASP: objects J1743 ($-0.055\pm0.003$~s~yr\textsuperscript{-1},
significant at 21~$\sigma$), J1331
($-0.075\pm0.013$~s~yr\textsuperscript{-1}, significant at 5~$\sigma$)
and J2344 ($-0.31\pm0.02$~s~yr\textsuperscript{-1}, significant at
16~$\sigma$).  Figs.~\ref{star17lc}--\ref{star44oc} show the folded
light curves and O$-$C diagrams for these three objects, while
Fig.~\ref{star13oc} illustrates a non-significant result (object
J0509) for comparison.  Given the method of period determination, the
choice of a precise zero point for the ephemerides of these three
objects is somewhat arbitrary.  However, using the zero points
assigned by the program, and the optimum periods for the whole
observation set, we have approximate eclipse times:
\begin{description}
\item[J1743] HJD $2454997.688926+0.258108E-4.5\times10^{-10}E^2$
\item[J1331] HJD $2454976.358606+0.218014E-5.2\times10^{-10}E^2$
\item[J2344] HJD $2454417.284057+0.213676E-2.1\times10^{-9}E^2$.
\end{description}

\section{Discussion}

We may note that period changes have been observed in many EBs
(\citealt{kim} claim that about 46\% of the 1140 EBs in
\citeauthor{kreiner}'s \citeyearpar{kreiner} collection of O$-$C
diagrams show at least some evidence of period change).  Moreover, in
short-period W\,UMa-type systems, period increases and decreases have
been found to be of similar frequency, with the majority having
$|\dot{P}| \leq 0.02$~s~yr\textsuperscript{-1} and none exceeding
about $\pm0.4$~s~yr\textsuperscript{-1} (\citealt{kubiak}), which
would make our 3 main examples of period change notably rapid, but not
exceptionally so.  Therefore, we do not suggest that our evidence for
period change is in itself especially surprising, nor would we argue
for period \emph{decreases} being generally dominant in this stellar
population, on the basis of such a small sample.  Rather, the
particular interest of (relatively rapid) period decrease in the three
objects J1743, J1331 and J2344 is that they are already near or below
the short-period limit for W\,UMa objects of 0.22~d, and have
timescales $P/\dot{P}$ between $4\times10^5$ and $6\times10^4$ years.
The underlying processes which may be operating in these objects
therefore merit further consideration.

To investigate possible causes of the period decreases apparently
observed for these three objects, the eclipsing binary modelling
software PHOEBE \citep{prsa}, built upon the code of \citet{wildev},
was used to estimate component masses, radii and other system
parameters.  In the absence of radial velocity measurements, values
for semimajor axis $a$ could not be determined, so best-fit parameter
combinations were found for a range of physically-plausible values of
$a$.  Several external constraints were applied: from general light
curve morphology, such as continuous light variation, the systems were
taken to be W\,UMa-type contact binaries (overcontact in PHOEBE's
terms) i.e. with filling factors in (0,1].  The values for $M_1$ were
required to be approximately consistent with system temperatures as
indicated by spectral types, themselves estimated from $V-J$ and $V-K$
colours using 2MASS data (J1743 was taken to be approximately type K2,
with temperature $\sim$~4600~K; J1331 type G9,
$T_{\mathrm{eff}}\sim$~5000~K; J2344 type K9,
$T_{\mathrm{eff}}\sim$~3700~K).  Mass ratios $q$ were taken to be in
the range 0.08--0.8, as typical for W\,UMa systems \citep{hilditch},
subject to the hydrogen-burning requirement for $M_2$ to be at least
$0.08M_{\sun}$.  Maximum and minimum values of $a$ could then be
derived from the possible radii of the (contact) component stars.  The
resulting best-fit masses and radii (found to 2~s.f. by minimizing
$\chi^2$ values) are given in Table~\ref{3model}.  Object J1331 could
not be as well-fitted as the other two objects, by any combination of
$a$, $q$, inclination $i$ and Kopal potential $\Omega$, because its
maxima differ significantly in height, probably due to spots on the
stellar surfaces.  However, spots were not modelled here, in the
absence of direct evidence.

Period decrease might be caused by magnetic braking, removing angular momentum
from the system.  The plausibility of this was assessed using
\citeauthor{bradguin}'s equation \citeyearpar{bradguin}:
\begin{equation}
\dot{P}\approx-1.1\times10^{-8}q^{-1}(1+q)^2(M_1+M_2)^{-\frac{5}{3}}k^2(M_1R_1^4-M_2R_2^4)P^{-\frac{7}{3}}
\end{equation}
Although this applies properly only to detached systems,
\citeauthor{bradguin} suggest that magnetic field strengths would be
weaker in contact binaries.  So we may at least use their formula to
estimate an upper limit for the effect of magnetic braking on a
contact system.  Therefore, taking the gyration constant $k^2$ as 0.1
(typical for main sequence stars, \citealt{bradguin}), the value of
the RHS was evaluated for the various system parameter combinations
and compared with $\mathrm{d}P/\mathrm{d}t$ for the three
stars.  The results, given in Table~\ref{3model}, are between about
20\% and 1\% of the observed quantities.  Other estimates of the
effect of magnetic braking were made from equations for $\dot{J}$ from
\citet{rap} and \citet{hur}, as calibrated by \citet{davis} by using
the angular momentum loss rate at the upper edge of the cataclysmic
variable period gap (the binary systems were approximated as single,
fully-convective stars in order to use these equations).  This
produced significantly smaller estimates of expected period change:
between 3 and 4 orders of magnitude too small.  This suggests that
magnetic braking is not the main cause of period decrease in any of
the three systems, though the difference of 1 to 2 orders of magnitude
in the estimates produced by the different equations also implies
significant limitations in our ability to quantify magnetic braking.

Another possibility is angular momentum loss due to gravitational wave
radiation (GWR).  Combining expressions from \citet{kolb} and \citet{hilditch}:
\begin{equation}
\frac{\dot{P}}{P}=3\frac{\dot{J}}{J}=-1.27\times10^{-8}\mathrm{yr}^{-1}\frac{M_1M_2}{(M_1+M_2)^{\frac{1}{3}}M_{\sun}^{\frac{5}{3}}}\left(\frac{P_{\mathrm{orb}}}{\mathrm{h}}\right)^{-\frac{8}{3}}
\end{equation}
the expected $\dot{P}/P$ was calculated for each parameter
combination (Table~\ref{3model}).  However, these quantities are about
6 orders of magnitude smaller than the observed $\dot{P}/P$
values.  GWR can be responsible for only a tiny fraction of the period
decreases observed here.

\begin{table*}
\caption{Model parameters and theoretical period/mass changes for three objects.}
\label{3model}
\centering
\begin{tabular}{l l l l l l l l l l l}
\hline\hline
Star & $a$ & $q$ & $M_1$ & $M_2$ & $R_1$ & $R_2$ & Magnetic & GWR & Non-con- & Conservative\\
Short & & & & & & & Braking & $\frac{\dot{P}}{P}$ & servative & mass transfer\\
Name & ($R_{\sun}$) & ($M_2$/$M_1$) & ($M_{\sun}$) & ($M_{\sun}$) & ($R_{\sun}$) & ($R_{\sun}$) & $\frac{\dot{P}}{P}$ (yr\textsuperscript{-1}) &
(yr\textsuperscript{-1}) & $\dot{M_1}$ ($M_{\sun}$~yr\textsuperscript{-1}) & $\dot{M_1}$ ($M_{\sun}$~yr\textsuperscript{-1})\\
\hline
J1743 & 1.5 & 0.32 & 0.52 & 0.17 & 0.74 & 0.45 & $-1.7\times10^{-7}$ & $-2.9\times10^{-11}$ & $-5.1\times10^{-7}$ & $-2.0\times10^{-7}$\\
 & 1.6 & 0.24 & 0.67 & 0.16 & 0.88 & 0.49 & $-3.5\times10^{-7}$ & $-3.4\times10^{-11}$ & $-5.9\times10^{-7}$ & $-1.7\times10^{-7}$\\
 & 1.7 & 0.23 & 0.81 & 0.18 & 0.94 & 0.52 & $-4.4\times10^{-7}$ & $-4.4\times10^{-11}$ & $-7.0\times10^{-7}$ & $-1.9\times10^{-7}$\\
 & 1.8 & 0.31 & 0.90 & 0.28 & 0.91 & 0.55 & $-2.7\times10^{-7}$ & $-7.0\times10^{-11}$ & $-8.8\times10^{-7}$ & $-3.3\times10^{-7}$\\
\hline
J1331 & 1.4 & 0.12 & 0.69 & 0.09 & 0.79 & 0.31 & $-7.5\times10^{-7}$ & $-3.0\times10^{-11}$ & $-8.0\times10^{-7}$ & $-1.3\times10^{-7}$\\
 & 1.5 & 0.20 & 0.80 & 0.16 & 0.79 & 0.38 & $-4.3\times10^{-7}$ & $-5.9\times10^{-11}$ & $-1.1\times10^{-6}$ & $-2.6\times10^{-7}$\\
 & 1.6 & 0.41 & 0.82 & 0.34 & 0.74 & 0.49 & $-1.7\times10^{-7}$ & $-1.2\times10^{-10}$ & $-1.5\times10^{-6}$ & $-7.6\times10^{-7}$\\
 & 1.7 & 0.65 & 0.84 & 0.55 & 0.71 & 0.58 & $-1.2\times10^{-7}$ & $-1.9\times10^{-10}$ & $-1.9\times10^{-6}$ & $-2.1\times10^{-6}$\\
\hline
J2344 & 1.3 & 0.14 & 0.57 & 0.08 & 0.72 & 0.30 & $-5.3\times10^{-7}$ & $-2.6\times10^{-11}$ & $-2.9\times10^{-6}$ & $-5.4\times10^{-7}$\\
 & 1.4 & 0.45 & 0.56 & 0.25 & 0.63 & 0.44 & $-1.3\times10^{-7}$ & $-7.3\times10^{-11}$ & $-4.5\times10^{-6}$ & $-2.6\times10^{-6}$\\
\hline
\end{tabular}
\end{table*}

This leaves mass transfer from $M_1$ to $M_2$ and/or mass (and hence
angular momentum) loss from the system, from other mechanisms, as
plausible causes.  Using equations as given by \citeauthor{hilditch}:
\begin{equation}
\frac{\dot{P}}{P}=\frac{3\dot{M_1}(M_1-M_2)}{M_1M_2}
\end{equation}
\begin{equation}
\frac{\dot{P}}{P}=3\dot{M_1}\left[\frac{(M_1+M_2)}{M_1M_2}\frac{d^2}{a^2}-\frac{M_2}{M_1(M_1+M_2)}\right]
\end{equation}
(where $d$ is the distance from the binary centre of mass to Lagrange
point $L_2$) for conservative and non-conservative mass loss
respectively, we can calculate the necessary values of $\dot{M_1}$ to
explain the observed $\dot{P}/P$ values (Table~\ref{3model}).
These are of the order of
$10^{-6}$-$10^{-7}M_{\sun}$~yr\textsuperscript{-1}, which we may note
is similar to the mass loss rate calculated by \citeauthor{hilditch}
for contact binary SV Cen, to explain its precipitous observed rate of
period decrease.  He argues that mass may be ejected during a rapid
phase of Roche-lobe overflow (or more generally, in a contact system)
through $L_2$, contributing to period decrease.  Possible mechanisms
for unstable mass transfer/loss leading to rapid period decrease in
contact or near-contact binaries are also discussed in \citet{rasio},
\citet{tylenda} and \citet{jiang}.  This explanation is left as the
most plausible for the observed period decreases in the objects J1743,
J1331 and J2344, with further small contributions from magnetic
braking and GWR.

What future, then, might we envisage for these three systems?  Since
their primary masses and mass ratios cannot currently be determined
with any precision, it is possible that they are subject to
\citeauthor{jiang}'s low mass limit \citeyearpar{jiang}, leading to
unstable mass transfer, or to \citeauthor{rasio}'s tidal instabilities
\citeyearpar{rasio}, and will undergo merger on a relatively short
timescale.  \citeauthor{rasio} proposes the orbital decay time
$t_{\mathrm{D}}\sim10^3$--$10^4$ years for an unstable W\,UMa system.
Another indication of the possible timescale for merger is given by
\citeauthor{tylenda}'s study \citeyearpar{tylenda} of the decaying
period of the contact binary progenitor of the V1309\,Sco outburst.
Using their exponential model for period decay, one can determine that
their object would have been decreasing in period at a rate of
0.3~s~yr\textsuperscript{-1} about 130 years before the observed
outburst (and presumed merger).  Since object J2344 appears to be
currently undergoing period decrease at approximately this rate, and
is already below the short-period limit, it is perhaps not
inconceivable that it might merge on a timescale of centuries or even
decades from now (assuming, of course, that its period decrease is
maintained).  Whatever their ultimate fate, these may be rare examples
of objects caught in a brief transitional stage between stable states.

\section{Conclusions}

Our study of the periods of 53 W\,UMa candidate binary stars, observed
with SuperWASP, confirmed that they are all very close to the
short-period limit, and thus constitute a useful sample of sources for
investigating the causes of this limit.  In three of the objects,
period decrease significant at 5~$\sigma$ or more was found during
their time of observation; the remaining objects' period changes were
consistent with zero change at 3~$\sigma$.  Modelling estimates for
possible system parameters indicated that neither magnetic braking nor
GWR are likely to be the primary cause of these period decreases.  The
rates of change observed in these three objects can best be explained
by unstable mass transfer from primary to secondary components, and/or
mass and angular momentum loss from the systems, which could lead to
merger on a relatively short timescale if the periods continue to
decrease.

These are potentially unusual transitional objects, capable of
shedding light on the evolution of EBs and the causes of the
short-period limit of W\,UMa binaries.  As such they would repay
further study and additional observations.  In future we hope to
obtain spectroscopic measurements of these systems, permitting more
precise stellar modelling and parameter determination.  We also intend
to run the programs developed here on other SuperWASP variable
objects, to search for further interesting period variation.

\begin{acknowledgements}
The WASP project is funded and operated by Queen's University Belfast,
the Universities of Keele, St. Andrews and Leicester, the Open
University, the Isaac Newton Group, the Instituto de Astrofisica de
Canarias, the South African Astronomical Observatory and by STFC.
This work was supported by the Science and Technology Funding Council
and the Open University.  We would like to thank the referee
S.~Rucinski for his constructive advice which has improved this paper.
\end{acknowledgements}

\bibliographystyle{aa}
\bibliography{reflist}

\end{document}